\documentclass[superscriptaddress,twocolumn]{revtex4}
\usepackage{amssymb}
\usepackage[tbtags]{amsmath}
\usepackage{graphicx}
\usepackage{epsfig,graphicx,times}

\begin{document}

\title{Generation and control of Greenberger-Horne-Zeilinger entanglement in superconducting circuits}
\author{L.F. Wei}
\affiliation{Frontier Research System, The Institute of Physical
and Chemical Research (RIKEN), Wako-shi, Saitama, 351-0198, Japan}
\affiliation{IQOQI, Department of Physics, Shanghai Jiaotong
University, Shanghai 200030, China}
\author{Yu-xi Liu}
\affiliation{Frontier Research System, The Institute of Physical and Chemical Research
(RIKEN), Wako-shi, Saitama, 351-0198, Japan}
\author{Franco Nori}
\affiliation{Frontier Research System, The Institute of Physical and Chemical Research
(RIKEN), Wako-shi, Saitama, 351-0198, Japan}
\affiliation{Center for Theoretical Physics, Physics Department, CSCS, The University of
Michigan, Ann Arbor, Michigan 48109-1040, USA}
\date{\today }

\begin{abstract}

Going beyond the entanglement of microscopic objects (such as
photons, spins, and ions), here we propose an efficient approach
to produce and control the quantum entanglement of three {\it
macroscopic} coupled superconducting qubits. By conditionally
rotating, one by one, selected Josephson charge qubits, we show
that their Greenberger-Horne-Zeilinger (GHZ) entangled states can
be deterministically generated. The existence of GHZ correlations
between these qubits could be experimentally demonstrated by
effective single-qubit operations followed by high-fidelity
single-shot readouts. The possibility of using the prepared GHZ
correlations to test the macroscopic conflict between the
noncommutativity of quantum mechanics and the commutativity of
classical physics is also discussed.

\vspace{0.3cm} PACS number(s): 03.65.Ud, 03.67.Lx, 85.35.Ds
\end{abstract}

\maketitle

{\it Introduction.---}Entanglement is one of the most essential
features of quantum mechanics and has no analogue in classical
physics. Mathematically, it means that the wave function of a
system composed of many particles cannot be separated into
independent wave functions, one for each particle. Physically,
entangled particles can display remarkable and counter-intuitive
quantum effects. For example, a measurement made on one particle
collapses the total wavefunction and thus instantaneously
determines the states of the other particles, even if they are far
apart.

The existence of entanglement has been experimentally
demonstrated~\cite{asp82} with, e.g., two photons separated far
apart (e.g., up to $500$ m) and two {\it closely-spaced} trapped
ions (e.g., {\it separated a few micrometers apart}). The obvious
violation of Bell's inequality in these two-qubit experiments {\it
statistically} verifies the conflict between the locality of
classical physics and the non-locality of quantum mechanics. Only
recently, the experimental study of entanglement has been
successfully extended to a system composed of more than two
qubits. For example, three-photon Greenberger-Horne-Zeilinger
(GHZ) entangled states~\cite{Bouwmeester00} have been
demonstrated, and then used to test the conflict between classical
local-realism and quantum non-locality using {\it definite}
predictions~\cite{GHZ90}, rather than the {\it statistical} ones
based on Bell's inequalities. Yet, besides the problem of detector
efficiency, the expected GHZ state in optical experiments could
not be deterministically prepared~\cite{Bouwmeester00} because: i)
each entangled photon-pair was generated in a small subset of all
pairs created in certain spontaneous processes, and ii) the
nondeterministic detection of a trigger photon among two pairs of
entangled photons was required.

Instead of fast-escaping photons, massive or macroscopic quantum
systems~\cite{Liang04} have also been extensively studied to
realize controllable multipartite quantum entanglement. The
three-qubit entanglement of {\it microscopic} Rydberg
atoms~\cite{Raimond00} and trapped-ions~\cite{Science04-wineland}
was prepared experimentally. Moreover, the GHZ state of massive
{\it macroscopic} ``particles" has also been demonstrated in
liquid NMR~\cite{GHZ-NMR-98}. However, the existence of nonlocal
correlations in these ``particles" cannot be settled, as the
correlated information between them will be completely mangled in
their readouts of {\it ensemble averages}.

Superconducting qubits~\cite{today05} provides an attractive
platform to control the genuine (rather than ensemble-pseudo-pure)
macroscopic quantum state. The sizes of the present ``particles",
e.g., Cooper-pairs boxes, and the distance between them are
typically on the order of microns. If the interbit couplings are
{\it switchable}, then
methods~\cite{Raimond00,Science04-wineland,Bouwmeester00}, working
well in photon- and trapped-ion systems, could be
applied~\cite{Han04} to generate and verify the GHZ entanglement
between the Josephson qubits. However, in all published (so far)
experiments the interactions between Josephson
qubits~\cite{today05} are {\it fixed} (either capacitively or
inductively), and thus the usually required single-qubit gates
cannot, in principle, be {\it strictly} implemented.

For the currently-existing experimental circuits with {\it
always-on coupling}, here we propose an effective approach to
deterministically generate three-qubit GHZ states by conditionally
rotating the selected qubits one by one. The existence of the
desirable GHZ entanglement is then reliably verified by using {\it
effective} single-qubit operations. The prepared GHZ entanglement
should allow to test quantum nonlocality by definite predictions
at a macroscopic level.

{\it Preparation of GHZ states.---} We consider the three-qubit
circuit sketched in Fig.~1; that is, only adding one qubit to the
experimentally-existing one~\cite{Pashkin03}. Three
superconducting-quantum-interference-device (SQUID) loops with
controllable Josephson energies produce three Josephson qubits,
fabricated a small distance apart (e.g., up to \textit{a few
micrometers}~\cite{Pashkin03}, as the case of entangled trapped
ions in Ref.~\cite{Science04-wineland}) and coupled via the
capacitances $C_{12}$ and $C_{23}$. The dynamics of the system can
be effectively restricted to the subspace spanned by the
computational basis, and be thus described by the following
simplified Hamiltonian
\begin{equation}
\hat{H}=\frac{1}{2}\sum_{j=1}^3\left[E_{C}^{(j)}\sigma
_{z}^{(j)}-E_{J}^{(j)}\sigma
_{x}^{(j)}\right]+\sum_{j=1}^2K_{j,j+1}\,\sigma _{z}^{(j)}\sigma
_{z}^{(j+1)}.
\end{equation}
Here,
$E_{C}^{(j)}$$=$$2e^2[\tilde{C}_{\Sigma_j}^{-1}(2n_{g_j}-1)$$+$$\sum_{k\neq
j}\tilde{C}_{j,k}^{-1}(2n_{g_k}-1)]$\,with
$n_{g_j}$$=$$C_{g_{j}}V_{j}/(2e)\sim 0.5$, is the effective
charging energy of the $j$th qubit, whose effective Josephson
energy is $E_{J}^{(j)}$$=$$2\varepsilon _{J}^{(j)}\cos( \pi \Phi
_{j}/\Phi _{0})$ with $\varepsilon _{J}^{(j)}$ the Josephson
energy of the single-junction and $\Phi _{0}$ the flux quantum.
The effective coupling energy between the $j$th qubit and the
$(j+1)$th one is $K_{j,j+1}$$=$$e^2\tilde{C}_{j,j+1}^{-1}$. Above,
$C_{\Sigma _{j}}$ is the sum of all capacitances connected to the
$j$th box, and other effective capacitances are defined by
$\tilde{C}_{\Sigma_1}=C_{\Sigma_1}/(1+C^2_{12}C_{\Sigma_3}/\tilde{C})$,\,
$\tilde{C}_{\Sigma_2}=\tilde{C}/(C_{\Sigma_1}C_{\Sigma_3})$,\,
$\tilde{C}_{\Sigma_3}=C_{\Sigma_3}/(1+C^2_{23}C_{\Sigma_1}/\tilde{C})$,\,
$\tilde{C}_{12}=\tilde{C}/(C_{\Sigma_3}C_{12})$,\,\
$\tilde{C}_{23}=\tilde{C}/(C_{\Sigma_1}C_{23})$,\,
$\tilde{C}_{13}=\tilde{C}/(C_{12}C_{23})$, with
$\tilde{C}$$=$$\prod_{j=1}^3C_{\Sigma_j}-C^2_{12}C_{\Sigma_3}-C^2_{23}C_{\Sigma_1}$.
The pesudospin operators are defined as $\sigma^{(j)}
_{z}$$=$$|0_j\rangle \langle 0_j|-|1_j\rangle \langle 1_j|$ and
$\sigma^{(j)}_{x}$$=$$|0_j\rangle \langle 1_j|+|1_j\rangle \langle
0_j|$. As the interbit-couplings are always on, the charge energy
$E_C^{(j)}$ of the $j$th qubit depends not only on the
gate-voltage applied to the $j$th qubit, but also on those applied
to the other two Cooper-pair boxes. Compared to the coupling
$K_{j,j+1}$ between nearest-neighboring qubits, the interaction of
two non-nearest-neighbor qubits (i.e., $K_{13}=e^2/\tilde{C}_{13}$
between the first and the third qubits), is very weak and thus has
been safely neglected~\cite{ref1}. Indeed, for the typical
experimental parameters: $C_J\sim 600\,aF,\,C_m\sim 30\,aF$, and
$C_g=0.6\,aF$ in Ref.~\cite{Pashkin03}, we have
$K_{13}/K_{12}$$=$$K_{13}/K_{23}< C_m/C_J = 0.05$ and
$K_{12}/2\varepsilon _{J}\sim 1/4$.

In principle, the coupled qubits cannot be individually
manipulated, as the nearest-neighbor capacitive couplings
$K_{j,j+1}$ are sufficiently strong. However, once the state of
the circuit is known, it is still possible to design certain
operations for only evolving the selected qubits and keeping the
remaining ones unchanged. Our preparation begins with the ground
state of the circuit $|\psi(0)\rangle=|000\rangle$, which can be
easily initialized. The expected GHZ state could be produced by
the following simple three-step pulse process~\cite{ref1}
\begin{eqnarray}
|\psi(0)\rangle&=&|000\rangle
\,\overset{\hat{U}_2(t_2)}{\longrightarrow}\,
\frac{1}{\sqrt{2}}(|000\rangle\pm i|010\rangle)\nonumber\\
&&\overset{\hat{U}_1(t_1)}{\longrightarrow}\,
\frac{1}{\sqrt{2}}(|000\rangle\mp|110\rangle)\nonumber\\
&&\overset{\hat{U}_3(t_3)}\longrightarrow\,
\frac{1}{\sqrt{2}}(|000\rangle\pm
i|111\rangle)=|\psi_{GHZ}^\pm\rangle.
\end{eqnarray}
The first evolution $\hat{U}_2(t_2)$, with
$\sin[E_J^{(2)}t_2/(2\hbar)]$$=$$\pm 1/\sqrt{2}$, is used to
superpose two logic states of the second qubit. This is achieved
by simply using a pulse that switches on the Josephson energy
$E_J^{(2)}$ and sets the charging energy
$E_{C}^{(2)}$$=$$-2(K_{12}+K_{23})$. The second (or third)
evolution $\hat{U}_1(t_1)$\, (or $\hat{U}_3(t_3)$) is achieved by
switching on the Josephson energy of the first (third) qubit and
setting its charging energy as $E_C^{(1)}$$=$$2K_{12}$ (or
$E_C^{(3)}=2K_{23}$). The corresponding duration is set to satisfy
the conditions\, $\sin[E_J^{(j)}t_j/(2\hbar)]$$=$$1$ and
$\cos(\gamma_jt_j/\hbar)$$=$$1$, with
$\gamma_j$$=$$\sqrt{(2\,K_{j2})^2+(E_J^{(j)}/2)^2}$, with $j=1,3$,
in order to conditionally flip the $j$th qubit; that is, flip it
if the second qubit is in the $|1\rangle$ state, and keep it
unchanged if the second qubit is in the $|0\rangle$ state.

The fidelity of the GHZ state prepared above can be experimentally
measured by quantum-state
tomography~\cite{zeilinger05,Science04-wineland,GHZ-NMR-98}.
However, it would be desirable to confirm the existence of a GHZ
state without using tomographic measurements on a sufficient
number of identically prepared copies. Optical
experiments~\cite{Bouwmeester00} have achieved this via
single-shot readout and we propose a superconducting-qubit analog
of this approach. The single-shot readout of a Josephson-charge
qubit has been experimentally demonstrated~\cite{NEC04} by using a
single-electron transistor (SET)~\cite{delsing01}. Before and
after the readout, the SET is physically decoupled from the qubit.
The GHZ state generated above implies that the three SETs, if they
are individually coupled to each one of the three Cooper-pair
boxes at the same time, will simultaneously either receive charge
signals or receive no signal. The former case indicates that the
circuit is in the state $|111\rangle$, while the latter one
corresponds to the state $|000\rangle$. However, the existence of
these two terms, $|111\rangle$ and $|000\rangle$, in these
single-shot readouts, is just a necessary but not yet sufficient
condition for demonstrating the GHZ entanglement. Indeed, a
statistical mixture of those two states may also give the same
measurement results. In order to confirm that the state (2), e.g.,
$|\psi_{GHZ}^+\rangle$, is indeed in a coherent superposition of
the states $|000\rangle$ and $|111\rangle$, we consider the
following operational sequence
\begin{eqnarray}
|\psi_{GHZ}^+\rangle&\overset{\tilde{U}_2}\longrightarrow&
\frac{1}{2}(|000\rangle-|101\rangle+i|010\rangle+i|111\rangle)\nonumber\\
&\overset{\hat{P}_2}\longrightarrow&
\frac{1}{\sqrt{2}}(|0_10_3\rangle+|1_11_3\rangle)\nonumber\\
&\overset{\tilde{U}_1\otimes\tilde{U}_3}\longrightarrow&
\frac{1}{\sqrt{2}}(|0_11_3\rangle+ |1_10_3\rangle),
\end{eqnarray}
which is similar to the verification of the optical GHZ
correlations~\cite{Bouwmeester00}. Above,
$\hat{P}_2=|1_2\rangle\langle 1_2|$ is a projective measurement of
the second qubit. The suffixes are introduced in the second and
third steps to denote the order of the qubits. When we finally
readout the first and third qubits at the same time, the
simultaneous absence of the terms $|0_10_3\rangle$ and
$|1_11_3\rangle$ due to destructive interference indicates the
desired coherent superposition of the terms in the prepared GHZ
state (2).  The question now is how to realize the required
single-qubit operations
$\tilde{U}_j=\exp[i\pi\sigma_x^{(j)}/4],\,j=1,2,3$, keeping the
remaining qubits unchanged, in this circuit with untunable
interbit interactions (like the currently available experimental
ones).

\begin{figure}[tbp]
\vspace{-0.6cm}
\includegraphics[width=18cm, height=10cm]{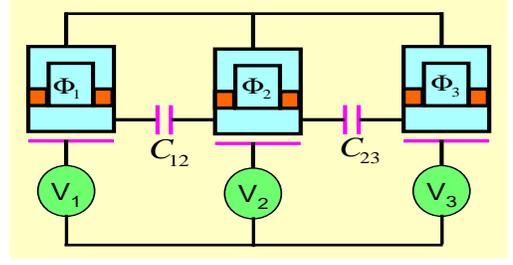}
\vspace{-6cm} \caption{(Color online) Three capacitively-coupled
SQUID-based charge qubits. The quantum states of three Cooper-pair
boxes (i.e., qubits) are manipulated by controlling the applied
gate-voltages $V_j\,(j=1,2,3),$ and external magnetic-fluxes
$\Phi_{j}$ (penetrating the SQUID loops). The circuit can be
generalized to include more qubits.}
\end{figure}

In order to effectively implement the single-qubit rotation
$\tilde{U}_2$ performed only on the second qubit, while keeping
the first and third qubits unchanged, we let the circuit evolve
under the Hamiltonian
$\hat{H}_2=-\varepsilon^{(2)}_J\sigma^{(2)}_x+K_{12}\,\sigma_z^{(1)}\sigma_z^{(2)}+K_{23}\,\sigma_z^{(2)}\sigma_z^{(3)}$,
by only switching on the Josephson energy of the second qubit,
e.g., $E^{(2)}_J=2\varepsilon_J^{(2)}$. Since
$\zeta_{12}=K_{12}/(2\varepsilon_J^{(2)})<
1,\,\zeta_{23}=K_{23}/(2\varepsilon_J^{(2)}) < 1$ (e.g., $\lesssim
1/4$ for the typical experimental parameters~\cite{Pashkin03}), we
can treat the second and third terms in $\hat{H}_2$ as
perturbations of the first one there. Indeed, neglecting
quantities smaller than the second-order
perturbations~\cite{wei05-PRB}, the Hamiltonian $\hat{H}_2$ can be
effectively approximated to~\cite{ref2}
\begin{equation}
\hat{H}_{\rm
eff}^{(2)}=-\varepsilon_J^{(2)}\left[1+2\,\zeta^2_{12}+2\,\zeta^2_{23}+4\,\zeta_{12}\,\zeta_{23}\,\sigma^{(1)}_z\sigma^{(3)}_z\right]\sigma
_{x}^{(2)}.
\end{equation}
In the state (2) the logic states of the first and third qubits
are always identical. Thus, by setting the corresponding duration
$\tau_2$ as
$\tau_2=\hbar\pi/\{4\varepsilon_J^{(2)}\left[1+2\zeta^2_{12}+2\zeta^2_{23}+4\zeta_{12}\zeta_{23}\right]\}$,
the required single-qubit operation
$\tilde{U}_2=\exp(-i\hat{H}_{\rm
eff}^{(2)}\tau_2/\hbar)=\exp(i\pi\sigma_x^{(2)}/4)$ could be
effectively performed on the second qubit in state (2). Similarly,
the Hamiltonian
$\hat{H}_{13}=\sum_{j=1,3}\{-\varepsilon^{(j)}_J\sigma^{(j)}_x+K_{j2}\sigma_z^{(j)}\sigma_z^{(2)}\}$,
induced by simultaneously switching on the Josephson energies of
the first and third qubits, can be effectively approximated to
\begin{eqnarray}
\hat{H}^{(13)}_{\rm
eff}=-\sum_{j=1,3}\varepsilon_J^{(j)}\left[1+2\,\zeta^2_{j2}\,\sigma_z^{(2)}\right]\sigma
_{x}^{(j)},
\end{eqnarray}
by neglecting the higher-order terms of
$\zeta_{j2}=K_{j2}/(2\varepsilon^{(j)}_J)<1$,\,with $j=1,3$. The
shifts of Josephson energies
$\Delta\tilde{E}_J^{(j)}=4\varepsilon_J^{(j)}\zeta^2_{j2}\,\sigma_z^{(2)}$
depend on the state of the second Cooper-pair box, which collapsed
into the state $|0\rangle$ after the projective measurement
$\hat{P}_2=|1_2\rangle\langle 1_2|$ (because such a measurement
tunnels the existing excess Cooper-pairs into the connected SET).
Thus, the effective Hamiltonian $\hat{H}_{\rm eff}^{(13)}$ yields
the evolution
$\hat{U}_{13}(\tau_{13})=\exp(-i\hat{H}^{(13)}_{\rm eff}\tau_{13}/\hbar)
=\prod_{j=1,3}\exp\left\{i\tau_{13}[\varepsilon_J^{(j)}(1+2\,\zeta^2_{j2})]\sigma^{(j)}_x/\hbar\right\}$.
Obviously, if the duration $\tau_{13}$ satisfies the condition
$\tau_{13}[\varepsilon_J^{(j)}(1+2\zeta^2_{j2})]/\hbar=\pi/4$,
then the required single-qubit operations
$\tilde{U}_{j}=\exp[i\pi\sigma_x^{(j)}/4]$ could be simultaneously
implemented.

{\it Possible application.---} The prepared GHZ state, e.g.,
$|\psi_{GHZ}^+\rangle$, should allow, at least in principle, to
test the {\it macroscopic} conflict between the noncommutativity
of quantum mechanics and the commutativity of classical physics by
definite predictions~\cite{GHZ90}. Using the EPR's reality
criterion, each observable corresponds to an ``element of reality"
(even if it is not measured). That is, the quantum operators
$\sigma^{(j)}_\alpha,\,(\alpha=x,y,z;\,j=1,2,3)$ are linked to the
classical numbers $m^{(j)}_\alpha$, which have the value $+1$ or
$-1$. The so called $\sigma^{(j)}_{\alpha}$-measurement is the
projection of the quantum state into one of the eigenstates of
$\sigma^{(j)}_\alpha$. The prepared GHZ state is the eigenstate of
the three operators:
$A_{yxx}=\sigma^{(1)}_y\sigma^{(2)}_x\sigma^{(3)}_x,
A_{xyx}=\sigma^{(1)}_x\sigma^{(2)}_y\sigma^{(3)}_x,$ and
$A_{xxy}=\sigma^{(1)}_x\sigma^{(2)}_x\sigma^{(3)}_y$, with a
common eigenvalue $+1$. Thus, {\it classical} reality implies that
\begin{eqnarray*}
1&=&(m^{(1)}_ym^{(2)}_xm^{(3)}_x)(m^{(1)}_xm^{(2)}_ym^{(3)}_x)
(m^{(1)}_xm^{(2)}_xm^{(3)}_y)\\
&=&m^{(1)}_ym^{(2)}_ym^{(3)}_y.
\end{eqnarray*}
The second formula indicates that, if we perform the
$\prod_{j=1}^3\sigma^{(j)}_y$-measurement (i.e., $yyy$-experiment)
on the state $|\psi^+_{GHZ}\rangle$, the eigenstate
$|\tilde{-}\rangle$ only shows in pairs. Here, $|\tilde{+}\rangle$
(or $|\tilde{-}\rangle$) denotes the eigenstate of the operator
$\sigma_y$ with eigenvalue $+1$ (or $-1$) and corresponds to the
classical number $m_y=+1$ (or $-1$). While, for this
$yyy$-experiment {\it quantum}-mechanics predicts that the state
$|\tilde{-}\rangle$ never shows simultaneously in pairs, because
the prepared GHZ state can be rewritten as $|\psi_{GHZ}^+\rangle=
(|\tilde{+}\tilde{+}\tilde{-}\rangle+|\tilde{+}\tilde{-}\tilde{+}\rangle
+|\tilde{-}\tilde{+}\tilde{+}\rangle+|\tilde{-}\tilde{-}\tilde{-}\rangle)/2$.
Obviously, this contradiction comes from the fact that the
observable $\sigma^{(j)}_x$ anti-commutes with the observable
$\sigma^{(j)}_y$ and the operator identity
$$
(\sigma^{(1)}_y\sigma^{(2)}_x\sigma^{(3)}_x)(\sigma^{(1)}_x\sigma^{(2)}_y\sigma^{(3)}_x)
(\sigma^{(1)}_x\sigma^{(2)}_x\sigma^{(3)}_y)=\,-\,\sigma^{(1)}_y\sigma^{(2)}_y\sigma^{(3)}_y,
$$
which is ``opposite" to its classical counterpart.

The protocol described above could be directly (e.g., for the
optical system~\cite{Bouwmeester00}) performed by reading out the
eigenstates of the operators $\sigma_x$ and $\sigma_y$,
respectively. However, in the present solid-state qubit, the
eigenstates of $\sigma_z$ are usually read out. Thus, additional
operations, e.g., the Hadamard transformation
$\hat{S}_x=(\sigma_z+\sigma_x)/\sqrt{2}$, and the unitary
transformation
$\hat{S}_y=[(1+i)\hat{I}+(1-i)\sum_{\alpha}\sigma_{\alpha}]/(2\sqrt{2})$,
are required to transform the eigenstates of $\sigma_x$ and
$\sigma_y$ to those of $\sigma_z$, respectively. These additional
single-qubit operations could be implemented by combining the
rotations of the selected qubit along the $x$-axis (by using the
effective Hamiltonian proposed above) and those along the $z$-axis
(by effectively refocusing the
fixed-interactions~\cite{wei05-PRB}).

{\it Conclusion and Discussions.---} The experimental realization
of our proposal for producing and testing GHZ correlations is
possible, although it may also face various technological
challenges, like other theoretical designs~\cite{Makhlin99} for
quantum engineering. Of course, the fabrication of the proposed
circuit is not difficult, as it only adds one qubit to
experimentally-existing superconducting
nanocircuits~\cite{Pashkin03}. Moreover, rapidly switching on/off
the Josephson energy, to realize the fast quantum manipulations,
is experimentally possible. In fact, assuming a SQUID loop size of
$10$ $(\mu $m$)^2,$ changing the flux by about a half of a flux
quantum in $10^{-10}$s, requires sweeping the magnetic field at a
rate of $10^{5}$ Tesla/s, almost reachable by current
techniques~\cite{Uwazumi02}.

Also, the prepared GHZ states are the eigenstates of the idle
circuit (i.e., no operations on it) without any charge- and
Josephson energies (by setting the controllable parameters as
$\Phi_j=\Phi_0/2$ and $n_{g_j}=0.5$) and thus are relatively
long-lived, at least theoretically. Indeed, the couplings
$\hat{V}=\sum_{j=1}^3\sigma_{z}^{(j)}\left(\sum_{k=1}^3\lambda_{jk}X_{k}\right)$
between the relevant baths and the circuit commute with the
non-fluctuating Hamiltonian of the idle circuit
$\hat{H}_0=\sum_{j=1,2}K_{j,j+1}\,\sigma_z^{(j)}\sigma_z^{(j+1)}$.
Here, $\lambda_{jk}$ equals to either $1$ for $j=k$ or
$\tilde{C}_{\Sigma_k}\tilde{C}_{jk}$ for $j\neq k$, and
$X_{k} =(e\,C_{g_k}/\tilde{C}_{\Sigma_k})\sum_{\omega
_{k}}(g_{\omega _{k}}^{\ast }\hat{a}_{\omega _{k}}^{\dag
}+g_{\omega _{k}}\hat{a}_{\omega _{k}})$
%
with $\hat{a}_{\omega _{k}},\hat{a}_{\omega _{k}}^{\dagger }$
being the Boson operators of the $k$th bath, and $g_{\omega _{k}}$
the coupling strength between the oscillator of frequency $\omega
_{k}$ and the non-dissipative system.
Thus, only pure dephasing, i.e., the zero frequency value of the
noise spectrum contributes to overall decoherence
rates~\cite{Markus03}. However, the working frequency of the
present circuit is always non-zero. This implies that the
lifetimes of the prepared GHZ correlations are still sufficiently
long, and thus various required quantum manipulations could still
be coherently implemented.

Perhaps, the biggest challenge comes from the fast single-shot
readouts~\cite{NEC04,Spiller03} of multi-qubits at the same time.
This is a common required task of almost all quantum algorithms
and an important goal for almost all physical realizations of
quantum computing. In order to avoid the crosstalk between qubits
during the readouts, the readout time $t_m$ should be ``much"
shorter than the characteristic time $t_c\sim \hbar/K_{j,j+1}$ of
communications.
This requirement has been achieved by the existing phase-qubit
circuits~\cite{martinis05}: $t_m\sim 1$\,ns, and $t_c\sim 4$\,ns
for the demonstrated coupling energy $K\sim 80$\,MHz. For the
existing charge-qubit circuits~\cite{Pashkin03}, where the
interbit coupling-energy $K\sim 3\,$GHz yields $t_c\sim 100$ ps,
the duration of the single-shot readout pulse should be not longer
than several tens of picosecond. Thus, the weaker interbit
coupling, e.g., lowered to hundreds of KHz, is required for the
current SET technique, whose response time is usually hundreds of
nanosecond~\cite{delsing01,NEC04}.

In summary, based on conditionally manipulating the selected
qubits, we have shown how to engineer the macroscopic quantum
entanglement of Josephson qubits with fixed-couplings. Our
proposal allows to deterministically prepare three-qubit GHZ
entangled states and allows a macroscopic test of the
contradiction between the noncommutativity of quantum mechanics
and the commutativity of classical physics.

This work was supported in part by the NSA and ARDA, under AFOSR
contract number F49620-02-1-0334, and by the NSF grant No.
EIA-0130383. We thank Dr. Xiang-bin Wang for useful discussions.


\begin{thebibliography}{99}
\bibitem{ref1}The operation $\hat{U}_2(t_2)$
is not influenced by $K_{13}$. The fidelity of the evolution by
the operation $\hat{U}_1(t_1)$ ($\hat{U}_3(t_3)$) achieves to
$99.95\%$ for the typical parameters:
$K_{13}/K_{12}=K_{13}/K_{23}=0.05$. If $K_{13}$ is considered, the
exact evolution could still be obtained by simply modifying the
gate volatge $V_1$ to satisfy the condition
$E_C^{(j)}=2K_{j2}-2K_{13}\,(j=1,3)$.
\bibitem{ref2} Compared to the exact evolution by $\hat{H}_2$, the
fidelity of the proposed evolution $\tilde{U}_2$ is $F=|\langle
\psi^+_{GHZ}|\tilde{U}_2^\dagger\exp(-i\hat{H}_2\tau_2)|\psi^+_{GHZ}\rangle|\sim
92.4\%\,\, (99.9\%)$ for the typical parameters
$K_{12}/\varepsilon_{J}^{(2)}=K_{23}/\varepsilon_{J}^{(2)}=0.5\,\,
(0.05)$.
\end{thebibliography}
\end{document}